\documentclass[pdftex,twocolumn,epjc3]{svjour3}          
\pdfoutput=1

\RequirePackage[T1]{fontenc}
\smartqed  
\RequirePackage{multirow}
\RequirePackage{graphicx}
\RequirePackage{mathptmx}      
\RequirePackage{amsmath}
\RequirePackage{caption}
\RequirePackage{subcaption}

\RequirePackage[numbers,sort&compress]{natbib}
\RequirePackage[colorlinks,citecolor=blue,urlcolor=blue,linkcolor=blue]{hyperref}

\graphicspath{ {./IMAGES/} }

\begin{document}
\title{Towards a Computer Vision Particle Flow
\thanksref{t1}}


\author{Francesco Armando Di Bello\thanksref{e5,addr3},
        Sanmay Ganguly\thanksref{e1,addr1},
        Eilam Gross\thanksref{addr1},
        Marumi Kado\thanksref{addr3,addr4},
        Michael Pitt\thanksref{addr2},
        Lorenzo Santi \thanksref{addr3},
        Jonathan Shlomi\thanksref{addr1}
}

\thankstext[$\star$]{t1}{Contact addresses}
\thankstext{e5}{e-mail: Francesco.Armando.DiBello@roma1.infn.it }
\thankstext{e1}{e-mail: sanmay.ganguly@weizmann.ac.il }

\institute{Weizmann Institute of Science, Rehovot 76100, Israel\label{addr1}
          \and
          CERN, CH\- 1211, Geneva 23, Switzerland\label{addr2}
          \and
          Universit\`a di Roma Sapienza, Piazza Aldo Moro, 2, 00185 Roma, Italy e INFN, Italy\label{addr3}
           \and
           Universit\' e Paris-Saclay, CNRS/IN2P3, IJCLab, 91405, Orsay, France\label{addr4}
}


\maketitle
\begin{abstract}


In High Energy Physics experiments Particle
Flow (PFlow) algorithms are designed to provide
an optimal reconstruction of the nature and
kinematic properties of the particles produced
within the detector acceptance during
collisions. At the heart of PFlow algorithms is
the ability to distinguish the calorimeter
energy deposits of neutral particles from those
of charged particles, using the complementary
measurements of charged particle tracking
devices, to provide a superior measurement of
the particle content and kinematics. In this
paper, a computer vision approach to this
fundamental aspect of PFlow algorithms, based on 
calorimeter images, is proposed. A
comparative study of the state of the art
deep learning techniques is performed. A
significantly improved reconstruction of the
neutral particle calorimeter energy deposits is
obtained in a context of large overlaps with
the deposits from charged particles. Calorimeter
images with augmented finer granularity are also obtained using super-resolution techniques.
\end{abstract}

\section{Introduction}
\label{sec:intro}

General-purpose high energy collider experiments are designed to measure both charged particle trajectories and \\
calorimeter clustered energy deposits. The charged particle tracks in a magnetic field and the topology of energy deposits in calorimeters provide most of the information necessary to reconstruct, identify and measure the energy of the particles that constitute the event, which for the most part are charged and neutral hadrons, photons, electrons, muons, and neutrinos. The latter escaping detection and are measured by the imbalance in momentum in electron-positron collision events or transverse momentum in hadron collision events. Other particles created during a high energy collision, having too short lifetimes to be directly detected in the experiment, need to be reconstructed from their decay products.

The goal of Particle Flow ({PFlow}) algorithms is to make optimal use of these complementary measurements to reconstruct the particle content and its energy response for the entire event. A precise reconstruction of the entire event is essential for the measurement of those particles, such as neutrinos, escaping detection, as well as the reconstruction of jets of particles originating from the fragmentation and hadronization of hard scattering partons. One challenging aspect of PFlow algorithms is to disentangle particles of different nature when they are close to one another and possibly overlap. The reconstruction performance in general and the performance of PFlow algorithms, in particular, will critically depend on the detector design specifications, as for instance, the size and magnetic field intensity of the tracking volume, the granularity of the calorimeters, and their energy resolution. The first PFlow algorithm was designed by the CELLO collaboration at PETRA~\cite{Behrend:1982gk}, where an optimal reconstruction of the event {\it "Energy Flow "} was measured by subtracting the expected energy loss from charged particles in the calorimeter, to estimate the {\it "neutral energy"} and its spatial distribution. This algorithm, developed in $e^+e^-$ collisions, was aimed at a precise measurement of the hadronic activity for the measurement of $\alpha_S$. Since then, PFlow algorithms relying on the parametrization of the expected charged energy deposits in calorimeters have been further developed at $e^+e^-$~\cite{Buskulic:1994wz} and $pp$~\cite{Aaboud:2017aca,Sirunyan:2017ulk} collider experiments. The success of these algorithms has been such that future $e^+e^-$ collider experiment projects are taking PFlow algorithms into account in the design of the projected detectors~\cite{Brient:2004zz,Thomson:2009rp,Ruan:2014paa,Gomez-Ceballos:2013zzn,CEPC-SPPCStudyGroup:2015csa}. 

In this paper, we explore the capabilities of computer vision 
algorithms, along with graph and deep set Neural Networks (NN),  to provide solutions to this complex question 
in a context of two overlapping particles, a charged and a neutral pion $\pi^{+}$ and $\pi^0$. This benchmark is highly representative in hadron collisions or jet environments in electron-positron collisions and served for instance as foundation to develop and tune the PFlow algorithm of the ATLAS experiment~\cite{Aaboud:2017aca}. It is the most common scenario in high energy hadron collisions where most of the event particle content originates from parton fragmentation and comprises typically of charged and neutral pions. The focus of this paper is the precise reconstruction of a "neutral energy" calorimeter image as a starting point for more elaborate particle flow and particle identification algorithms. In addition, images with an augmented output granularity are produced using super-resolution~\cite{2015arXiv150100092D} techniques. The ability to provide images with  finer granularity than the native segmentation of the detector is a very interesting possible intermediate step in particle identification and is also relevant in projecting the design granularity of future detectors.

Image-based convolutional NN have been used to study calorimeter shower and also used for hadronic jet tagging, as can be found Ref. \cite{Cogan:2014oua,
deOliveira:2015xxd,deOliveira:2018lqd,
Belayneh:2019vyx}. These approaches are based on uniform two or three-dimensional (2D or 3D) images. 
The more intricate case of varying granularity layered images has been recently addressed using graph NN in Ref.~\cite{Qasim:2019otl}. The hadronic showers with varying granularity layers are studied in the context of a future circular collider in Ref.~\cite{Aleksa:2019pvl}. 
Direct reconstruction of individual particles from detector hits has been investigated using segmentation of  particles through edge classifiers in the context of tracking~\cite{Farrell:2018cjr}, as well as direct segmentation and property determination using object condensation techniques~\cite{Kieseler:2020wcq}.

The approach adopted in this paper does not attempt the complete leap from the detector signals to fully identified high-level Particle Flow objects. It is a first intermediate step focusing on the reconstruction of a precise calorimeter image of the neutral particles in the event, using several NN architectures, including Convolution Neural Networks (ConvNet, UNet), Graph Neural Networks (Graph), and Deep Sets (DeepSet).
These calorimeter images are based on a detailed full Monte Carlo simulation, thus providing a precise benchmark to probe the performance of the proposed methods. 
To probe the performance of super resolution in a typical detector configuration, a native lower granularity is chosen in most layers of the calorimeters.


 \begin{figure*}
\begin{minipage}{\columnwidth}
\hspace{1.5cm}
\includegraphics[width=1.8\textwidth]{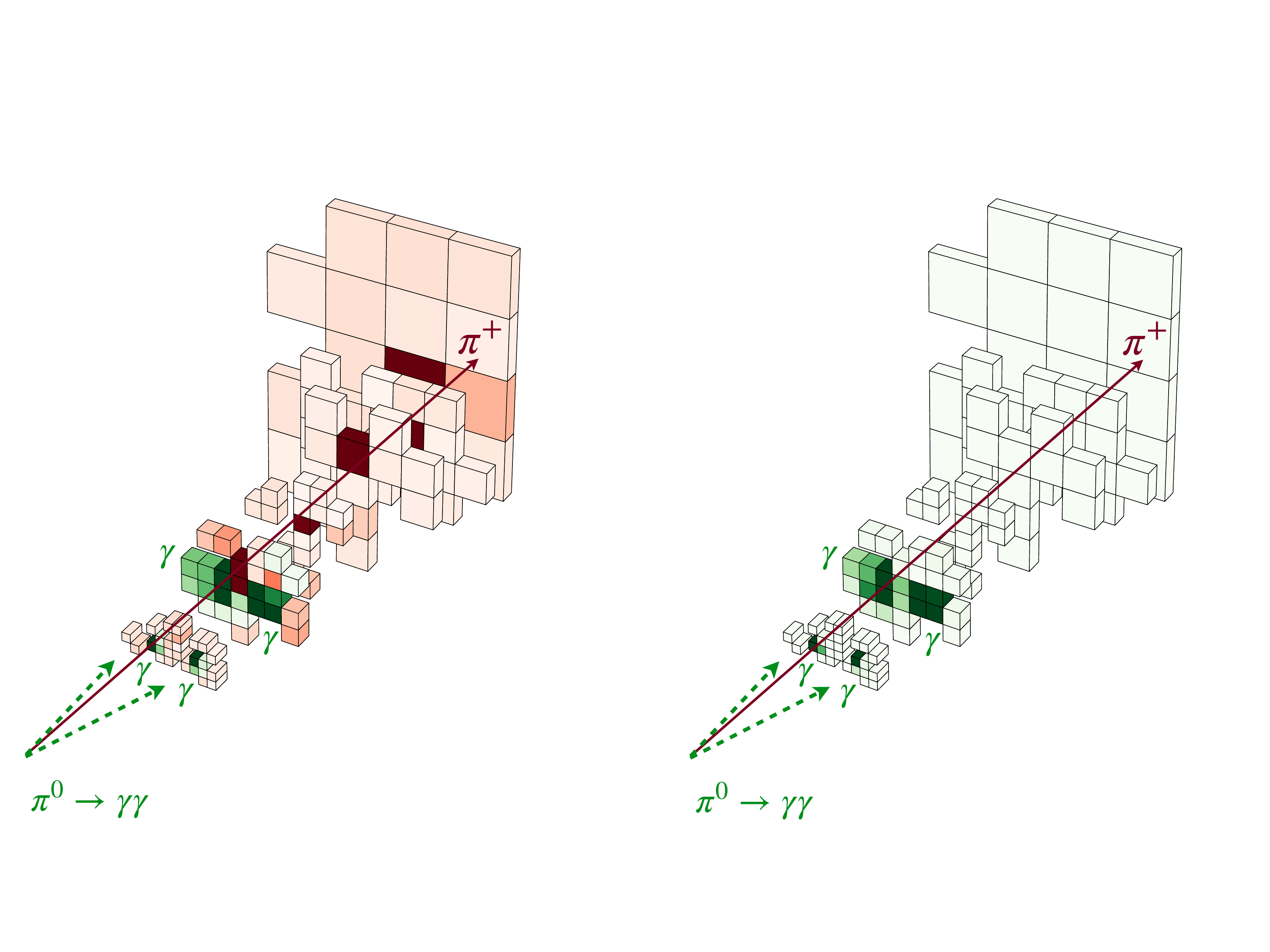}
 \end{minipage}
 \caption{A 3-D display the energy deposits of  $\pi^{+}$  and $\pi^{0}\rightarrow \gamma\gamma$ in the LG calorimeters (left) and  of $\pi^{0}$ only (right). The $\pi^{+}$ track and its extrapolation to the calorimeters is also  displayed. Cells, where the fraction of deposited energy from the $\pi^{+}$ is dominant are illustrated in red. The neutral energy deposits originating from the $\pi^{0}$ are otherwise illustrated in green. The extrapolation of the $\pi^{+}$ trajectory is also indicated.
 }
 \label{Fig:Regular_Res_EvDisplay_3D}
\end{figure*}

\begin{figure*}
\begin{minipage}{\columnwidth}
\includegraphics[width=2.0\textwidth]{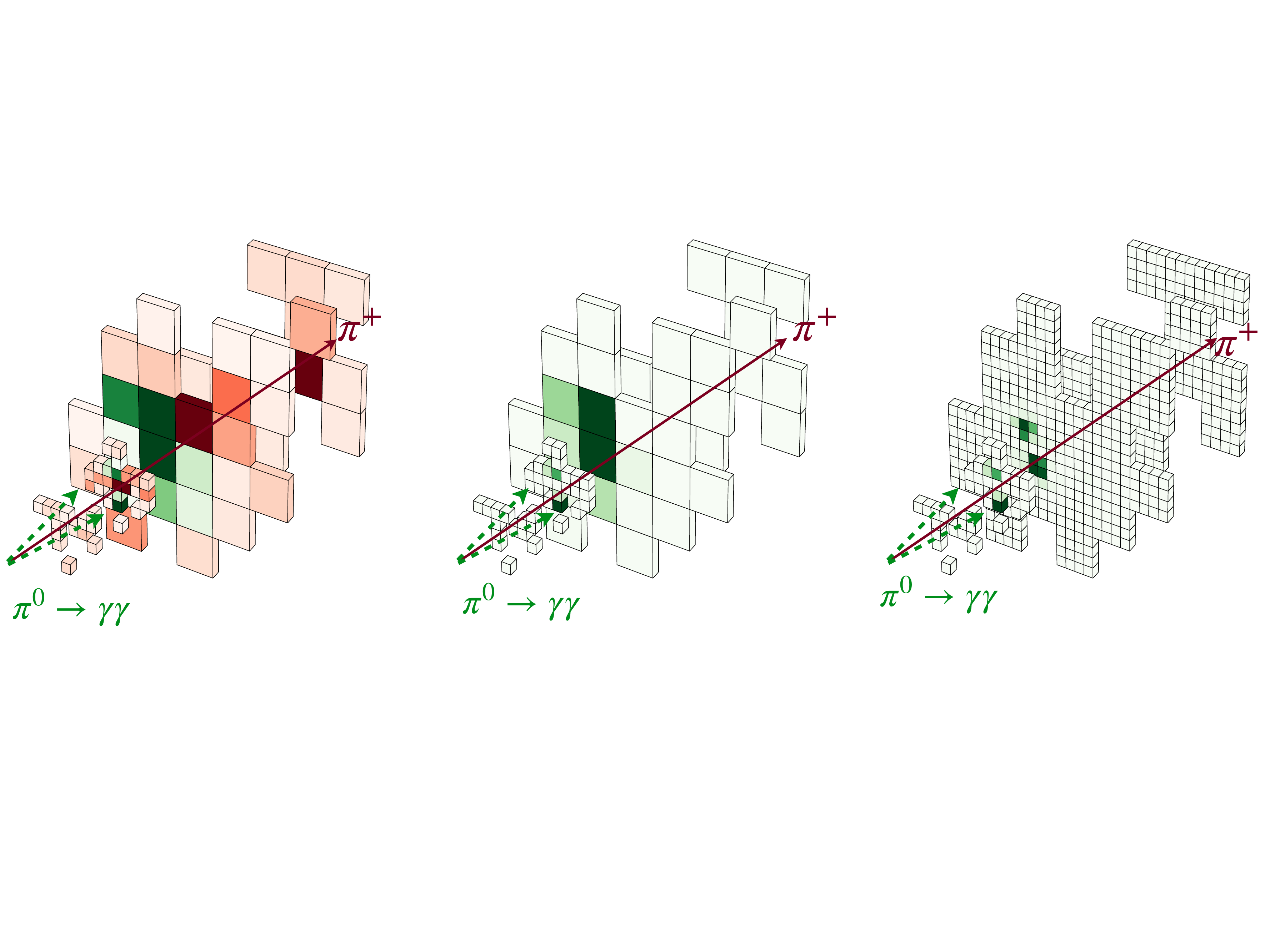}
 \end{minipage}
 \vspace{-3.5cm}
 \caption{A 3-D display the energy deposits of a $\pi^{+}$ and $\pi^{0}\rightarrow \gamma\gamma$, in the LG calorimeters (left), of the $\pi^{0}$ only (middle) and of
the same $\pi^{0}$ only shower as captured by a HG calorimeter layer of $32 \times 32$ granularity where the deposits from the two photons are resolved (right). The $\pi^{+}$ track and its extrapolation to the calorimeters are also  displayed.
 }
 \label{Fig:Super_Res_EvDisplay_3D}
\end{figure*}

\section{Detector Model and Simulation}
\label{sec:detector}

Two calorimeter layouts are considered in this paper, both based on the same underlying detector, but with different granularities. The two layouts will be referred to as High Granularity (HG) and Low Granularity (LG). The HG detector is used to study the separation of charged and neutral energy components while the LG detector is used as a benchmark for super-resolution applications. 

The experimental setup is composed of three sub-detectors. The electromagnetic calorimeter (ECAL) and hadronic \\ calorimeter (HCAL)
which are fully simulated, and a parameterized simulation of a charged particle tracker.
Each calorimetric sub-detector consists of three layers, 
with varying granularity, placed subsequently one after another. 

Both, the ECAL and the HCAL are sampling calorimeters where for simplicity and to produce a realistic shower development each layer is modeled as a homogeneous calorimeter using an equivalent molecule corresponding to a mixture of 
the absorber and scintillator materials. . 

The simulated energy deposits in each layer are then smeared to reproduce the corresponding sampling energy resolution. 
For the ECAL, the absorber material is lead and the active material is liquid argon, mixed with a mass proportion of $1.2:4.5$, 
yielding an equivalent radiation length of $X_0~=~3.9~cm$. 
For the HCAL, the absorber material used is iron with 
polyvinyl toluene plastic material used as a scintillating active material. These are 
combined with a mass proportion of 4.7:1.0 yielding an equivalent 
interaction length of $\lambda_{int}~=~17.4~cm$. The choices of the detector geometry, material and smearing parameters
are tuned to replicate single-particle energy responses similar to the one obtained with the ATLAS detector.

Overall the total length of the detector is 227.5~$cm$, including a $1~cm$ gap between the ECAL and HCAL blocks. The lateral profile of the calorimeters are squares of $125 \times 125$~$cm^2$. The main characteristics of the calorimeters are summarized in Table \ref{tab:detconfig}.

\begin{table}[ht]
\caption{Material budget of the ECAL and HCAL calorimeters as well as the corresponding equivalent radiation $X_0$ and interaction $\lambda_{int} $ lengths.}
\begin{center}
\begin{tabular}{c|c|c|l}
    \hline
    Detector &    Absorber &   Scintillator &  Subdetector (Legth)\\
    \hline
    ECAL             &    Lead        &   Liquid Argon & 
    ECAL1 (  {\bf \ \  3 $ \bf X_0$}  )\\
      &                 &        &         ECAL2 ( {\bf 16 $\bf X_0$} ) \\
      &                 &        &         ECAL3 ( {\ \  \bf 6 $\bf X_0$} ) \\
    \hline
    HCAL             &    Iron        &   {Plastic organic} & 
               HCAL1 (  {\bf 1.5 $ \bf \lambda_{int}$}  )\\
      &                 &        &         HCAL2 ( {\bf 4.1 $\bf \lambda_{int}$} ) \\
      &                 &        &         HCAL3 ( {\bf 1.8 $\bf \lambda_{int}$} ) \\
      \hline
\end{tabular}
\end{center}
\label{tab:detconfig}
\end{table}

The transverse granularities of each layer for both the HG and LG layouts are indicated in Table~\ref{tab:varres}.  An 
event display showing the deposited charged and neutral energy components for the HG detector
is shown in Figure \ref{Fig:Regular_Res_EvDisplay_3D}. A  similar event display for the LG detector is shown in Figure \ref{Fig:Super_Res_EvDisplay_3D}. A higher granularity detector composed of layers with 32$\times$32 resolution  is also shown on the same figure. It appears evident how the second layer of the LG detector is not able to separate between the two photon clusters in contrast to the 32$\times$32 layers.

\section{Simulated data}
\label{sec:dataset}

The electromagnetic showers of the photons from the subsequent decay of the neutral pion and the hadronic shower of the charged pion are simulated using GEANT~4~\cite{ALLISON2016186} using the detector layout described in Section~\ref{sec:detector}. Electronic noise in the calorimeter is also taken into account. 

To ensure significant overlap between the charged and neutral hadron, the polar angle of the $\pi^{+}$/$\pi^{0}$  momenta varies
randomly between $\pi / 100 $ to $3 \pi / 200$ radians, 
whereas the azimuthal angle varies  uniformly between 
$0$ and $2 \pi$ radians with a relative separation of $\pi / 60$ radians. 
The $\pi^{+}$ and $\pi^{0}$ are generated using the GEANT 
particle gun functionality. 
The  source of the gun is located $150~cm$ 
away w.r.t. the first ECAL layer.
To populate different parts of the detector, 
the initial location of the neutral and charged hadron in the event is  randomly chosen from the corner of a square at the source 
location with a length size equal to $20~cm$. 
Four sets of independent simulations are run with different 
energy ranges of:  $2-5$ GeV, $5-10$ GeV, $10-15$ GeV and $15-20$ GeV. 
The energy of the generated charged and neutral pions are 
randomly sampled from a uniform distribution bounded within these ranges, without any
correlation among the pions energies. 
The generation parameters of the particle gun ensure 
that a large proportion of detector cells have a 
significant amount of energy overlap, 
originating from the individual showers. The average fraction of neutral energy within groups of clustered cells, referred to as topoclusters (see Section \ref{sec:implementation}), is around 60\%.
The effect of electronic noise is emulated using 
gaussian distributions centered at zero with 
variables widths for different layers. The per cell levels of noise in each layer are given in Table \ref{tab:varres}.  For each cell in the event an energy is sampled from these distributions and added to the total energy. \\
To cross-check the effect of energy boundary transition,  a sample with $\pi^{+} \,, \pi^{0}$ energies randomly varied berween $2-20$ GeV  was also produced.

Finally, a track is formed by smearing the $\pi^{+}$ 
momentum by a resolution $\sigma(p)$, 
given by  $\frac{\sigma(p)}{p} = 5 \times 10^{-4} \times p~[GeV]$, and  keeping the original $\pi^{+}$ 
momentum direction unchanged.
The chosen momentum resolution of $\pi^{+}$ emulates the track resolution of the ATLAS tracking system and track reconstruction algorithms~\cite{Collaboration_2008}. The smearing of the track direction is neglected as it is expected to have sub-dominant effects to the results presented in this document.

\begin{table}[ht]
\caption{Transverse segmentations for both the HG and LG layouts (the total transverse dimension is $125 \times 125~cm^2$) of the ECAL and HCAL individual layers and the corresponding simulated electronic noise per cell for the HG detector are shown. The noise for the LG detector is appropriately scaled up by a conversion factor (cf) while transiting from HG to LG detector.}
\begin{center}
\begin{tabular}{c|c|c|c}
    \hline
    Detector Layer &    Res. (HG) & Res. (LG) &  Noise [MeV] (cf)\\
    \hline
    ECAL1        & $ 64 \times 64$ &  $ 32 \times 32$    & 13 ~(4)\\ \hline 
    ECAL2        & $ 32 \times 32$ &  $ 8 \times 8 $   & 34 ~(16)\\ \hline 
    ECAL3        & $ 32 \times 32$ &  $ 8 \times 8 $    & 17 ~(16)\\ \hline
    HCAL1        & $ 16 \times 16$ &  $ 8 \times 8 $    & 14 ~(4)\\ \hline 
    HCAL2        & $ 16 \times 16$ &  $ 8 \times 8 $    & 8 ~(4)\\ \hline 
    HCAL3        & $ 8 \times 8 $  &  $ 8 \times 8 $ &  14~~(1)\\ \hline 
\end{tabular}
\end{center}
\label{tab:varres}
\end{table}

\section{Deep neural network models} 
\label{sec:model}

The target of the NN models is to regress the per-cell neutral energy fraction using deep learning
methods to yield an accurate image of the neutral energy deposits. Two main approaches were investigated depending on the granularity of the target detector: a standard scenario where the granularity of the inputs and output images is unchanged, and a super-resolution scenario where the target detector features higher granularity layers compared to the input detector. For both, various state of the art NN architectures were implemented and compared.

For the standard scenario, the loss function is designed to regress the neutral energy fraction of each cell in the event, with a larger weight assigned to more energetic cells, to reduce the effect of noise and simultaneously enrich the performance of high energetic cells originating from the pions. The same loss function is used to train the different models and  defined on an event-basis as follows:

\begin{center}
$$L_{event} = \frac{1}{E_{tot}} \sum_c E_c (f^c_t-f^c_d)^2$$
\end{center}
where $E_{tot}$ is the total energy collected by the six calorimeter layers, $E_c$ is the total energy of a given cell indexed by $c$, $f^c_t$ and $f^c_d$ represent the target and predicted energy fractions.

Similarly to the previous case, the loss function used for super-resolution is built to regress the fraction of neutral energy of each  $super$-$cell$ ($f^{sc}$) with respect to the energy of the corresponding standard cell : 

\begin{center}
$$L^{\textrm{super-res}}_{event} =  \frac{1}{E_{tot}} \sum_{c} E_{c} \sum_{s=0}^{us^2} (f^{sc}_t-f^{sc}_d)^2$$
\end{center}
where $E_{c}$ is the total energy of the standard cell and $s$ is an index running over the super-cells  belonging to a standard cell $c$. Here $f^{sc}_t$ and $f^{sc}_d$ represents the target and predicted neutral energy fraction for the super-resoluted cell. The upper limit on the nested sum is the Up-Scale ($us$) factor used to define the high-resolution image. As an example, in this work we consider $us$ = 4 and thus each cell of the LG detector is up-scaled to 16 cells in the high-resolution image. 

For some of the NN models, a different loss function  was also studied and provided similar results:

\begin{center}
$$L^{2}_{event} = \frac{1}{N_{sc}} \sum_{sc} (e^{sc}_t-e^{sc}_d)^2$$
\end{center}
where $e^{sc}_t$ is the absolute value of the target neutral energy in the super-cell $sc$ and $e^{sc}_d$ is the corresponding predicted neutral energy. $N_{sc}$ is the total number of super-cells in an event that belongs to a topocluster. 

All the networks are trained with the Adam optimizer~\cite{kingma2014adam} and a
learning rate of $10^{-4}$. The training dataset consists of 
80,000 images while the validation dataset has 20,000 images. 
The performance of the trained models are evaluated on a test 
dataset consisting of 6000 images. The 
relative difference between the predicted neutral
energy and truth neutral energy in the 
test sample serves as the baseline metric.

In the following subsections, we describe the individual network architectures for standard resolution networks, where the input and output images have the same granularity, and for super-resolution networks where the output images have higher granularity compared to the inputs. 

\subsection{Networks used for standard resolution}
\subsubsection{Convolutional Network}
\label{sec:convnetmodel}

Calorimeter images from the total energy
deposit in the calorimeter cells in each layer are formed. The layers correspond to image channels (similarly to RGB in standard images),  each 
having a different resolution due to variable granularity of the 
calorimeter layers. To perform an image recognition analysis, each of the image channels (with a granularity given in Table~\ref{tab:varres}) is first mapped to a uniform resolution of $64 \times 64$ pixels. This mapping is performed by an individual NN block referred to as {\it "UpConv"} block. The mathematical operations of the UpConv block is described in Equation~ \ref{eqn:UpConvFlow}. 
The track is represented by the track layer, a single channel $64 \times 64$ pixel image. The image has only one non-zero pixel  
 which contains the value of the track momentum.
The location of 
this pixel in the image is determined from the  
position of the $\pi^{+}$ impact on the first ECAL layer surface. 
The track information is crucial in the design and performance 
of particle flow algorithm as it provides invaluable information to estimate the 
expected charged energy deposits as well as the  
position of the shower. 

The track layer is combined with the six Up-Convoluted calorimeter images to form a seven-layer image that is fed to a NN block consisting of convolutional layers, termed  {\it "ConvNet"} block. The output of 
the ConvNet block is a six images with a uniform granularity of $64 \times 64$ pixels.
Each image is then mapped to its corresponding native granularity through a down-convolutional learnable NN, denoted {\it "DownConv"} block. More details about the UpConv and DownConv blocks can be found in \ref{sec:upsample}. 

\subsubsection{UNet With ResNet }
\label{sec:unet}

An alternative to the image convolution based network
described above in section
\ref{sec:convnetmodel}, is the encoder-decoder network
(UNet)~\cite{ronneberger2015unet}. Similarly to the ConvNet architecture the UNet block is introduced after the UpConv blocks which produce 6 uniform $ 64 \times 64$ images forming, with the addition of the track image, the inputs to the UNet block. 
The encoder part
consists of four ResNet blocks (a 2D convolution
layer with a feed forward operation), followed by a 
bottleneck layer and four decoder blocks which 
perform the 2D transpose convolution to expand the
image size. There are also skip-connections between the output and input of encoder and decoder blocks with similar tensor shapes.
In our network, the output of the UNet 
block is a $6 \times 64 \times 64$ image which require to be mapped to the native resolution images through DownConv blocks.

\subsubsection{Graph Network}
\label{sec:graphnet}

The two networks described in subsections \ref{sec:convnetmodel} and \ref{sec:unet} are well suited for calorimeters with 
geometries with cell boundaries that can easily be mapped to common {\it "UpConv"} blocks. 
For calorimeters with irregular geometries, or where cell sizes across layers are not related through multiplicative factors, these networks are not ideal. Graphs do not have this limitation, as discussed in Ref. \cite{Shlomi:2020gdn}, and thus provide the most natural representation of the event. The energy deposits in the calorimeter cells along
with the tracks in the event have a natural representation
of a point cloud. Each calorimeter cell has a 3D 
position coordinate $(x, y, z)$ respectively. 
The $z$ coordinate of a calorimeter cell is defined
as the mid-point along $z$ axis, of the corresponding
calorimeter layer. 
The transverse location $(x_{tr}, y_{tr})$ where 
the track hits the ECAL1, denoting the point of entrance of the charged particle in the calorimeter volume where it will start developing its hadronic shower, is assigned to one node of the GNN.  

The $Graph$ $Network$ model is based on a Dynamic Graph CNN, similar to that used in Ref.~\cite{PhysRevD.101.056019}. 
A graph is formed dynamically at each iteration of the message passing using a KNN~\cite{macqueen1967}, with a default choice of $K = 10$. For the initial graph, each node has four embedding
features: three spatial coordinates and one
energy. In the chosen architecture, in addition to the spatial coordinates, the energy information is added. However spatial coordinates and energies are not directly combined. More details on graph network architecture can be found in \ref{sec:graphim}. 

\subsubsection{Deep Set Network}
\label{sec:deepset}

The point cloud description, as explained in 
\ref{sec:graphnet}, can be represented as a permutation equivariant (PE) set. Each element of
the set has four features (three coordinates and
one energy parameter).  
In the Deep Set (DS) formalism, described in Ref.~\cite{zaheer2018deep}, the forward passing through 
a DS layer is done through a PE operation : 
\begin{equation}
  \begin{aligned}
  f(\mathbf{x}) = \sigma\Big( \gamma~ \mathrm{I} ~\mathbf{x} - \lambda ~ mean(\mathbf{x} ) ~\mathrm{I}~ \Big) ~ ,
  \end{aligned}
  \label{eqn:DeepSetConv}
\end{equation}
where $\gamma$ and $\lambda$ are MLP's . 
$\sigma$ is the activation function which we choose to be a $\tanh$ function.

We design the DS layer to have
two parameters $P$ and $Q$ which 
are the dimensions of input and output features, respectively.

In our 
DS model, there are a total of four DS layers, followed 
by an MLP. The ($P ~\,, ~Q$) values for these four layers are (4, 6), (6, 12), (12,8), (8, 4), respectively. The output of each DS layer is concatenated
to one large vector which is then passed through the
final MLP block to get the target fractions on each 
node of the point cloud.

\subsection{Networks used for super-resolution}
\label{sec:superresmodels}

To solve the super-resolution task three approaches are used, based on graph networks, convolutional networks and a hybrid network composed of graph and convolutional networks.

\subsubsection{Pure graph network}
\label{sec:gnnonly}

The graph only network approach to the 
super-resolution problem is slightly different from that used for the regular energy flow
problem and requires a dual architecture with two main blocks connected through a broadcasting layer. The first block (GNN1), takes as an input the graph from the LG data and performs node and edge update operations~\cite{Shlomi:2020gdn,battaglia2018relational}. The broadcasting layer takes each node from the output of the GNN1 block and replicates it by the corresponding upsampling factor, from which a second block (GNN2) performs node and edge update operations. The output is then used to compare the loss function to the target graph, made of a higher number of nodes.

\subsubsection{Graph UNet network}
\label{sec:gnnunet}
The dynamic graph network and UNet are described in section \ref{sec:graphnet} and \ref{sec:unet} respectively. For the super-resolution task, we combine these two networks to form a hybrid network Graph-UNet. 

The graph network takes as an input the whole event graph and tries to predict the neutral energy per-cell of the topocluster for LG data. The UNet takes this prediction as input and upscales it via transpose convolution operation to match the granularity of the super-resolution image. An L2 loss function is then used to optimize the whole network and predict neutral energy for the super-resolution cells.

\subsubsection{Convolutional UNet network}
\label{sec:convunet}

The UNet network described in section \ref{sec:unet}, by construction, maps the input LG image to a high granularity image which is then scaled down through a DownConv block. Here we omit the DownConv block and directly try to regress the output of the UNet network to the target.  

\begin{figure*} 
\centering
\includegraphics[width=0.9\textwidth]{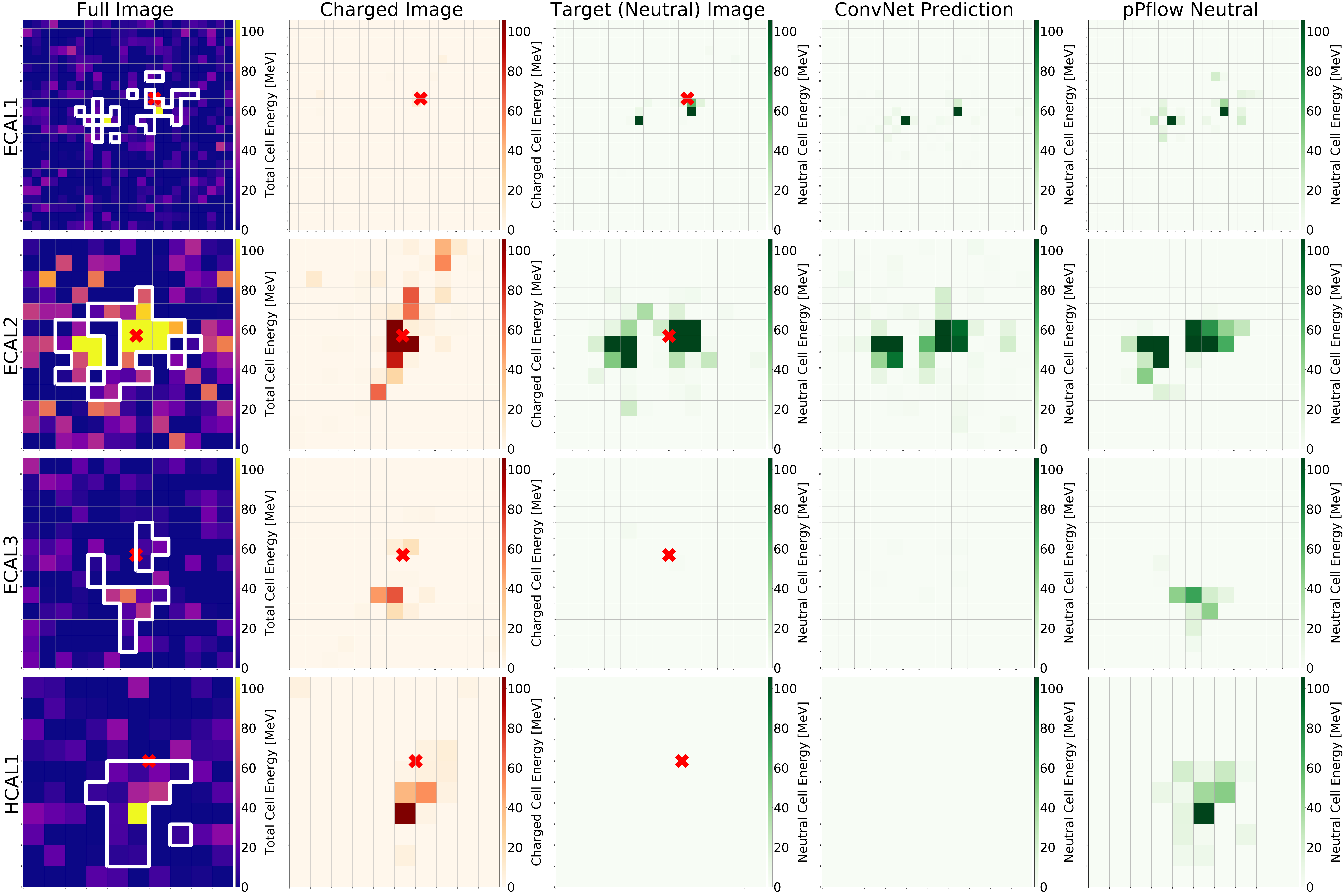}
\caption{An event display showing the 
individual components of charged and 
neutral energy per layer along with the 
noise distribution for the first four calorimeter layers. The 
 the extrapolated track position is shown by the 
 red cross for the first three columns.
The energy of the 
 initial $\pi^{+},\pi^{0}$ ranges 
 between 2-5 $GeV$. The total energy also includes noise contribution and topoclusters boundaries are outlined in white. The fourth and fifth columns 
 shows the predicted neutral energy distribution 
 from for the ConvNet and pPflow algorithms, respectively.
 In this example,
 it is seen that the electromagnetic 
 shower from 
 $\pi^{0} \rightarrow \gamma \gamma$ decay 
 has been correctly predicted by the trained
 model. The trained model also learns to suppress
 the noise pattern.}
\label{Fig:EventDisplayML}
\end{figure*}

\section{Parametric algorithm implementation}
\label{sec:implementation}
To quantify the performance of the designed NN algorithm, we compare the performance of the above mentioned NN algorithms to a traditional ATLAS like parametrized PFlow (pPflow) algorithm.

The pPflow algorithm is divided into two separate steps: (i)
the topocluster formation and (ii) the  
expected charged energy subtraction. The implementation of both steps is inspired by the PFlow algorithm currently used by 
the ATLAS experiment~\cite{Aaboud:2017aca}. 

\begin{figure*} 
\begin{minipage}{\columnwidth}
 \includegraphics[width=2\textwidth]{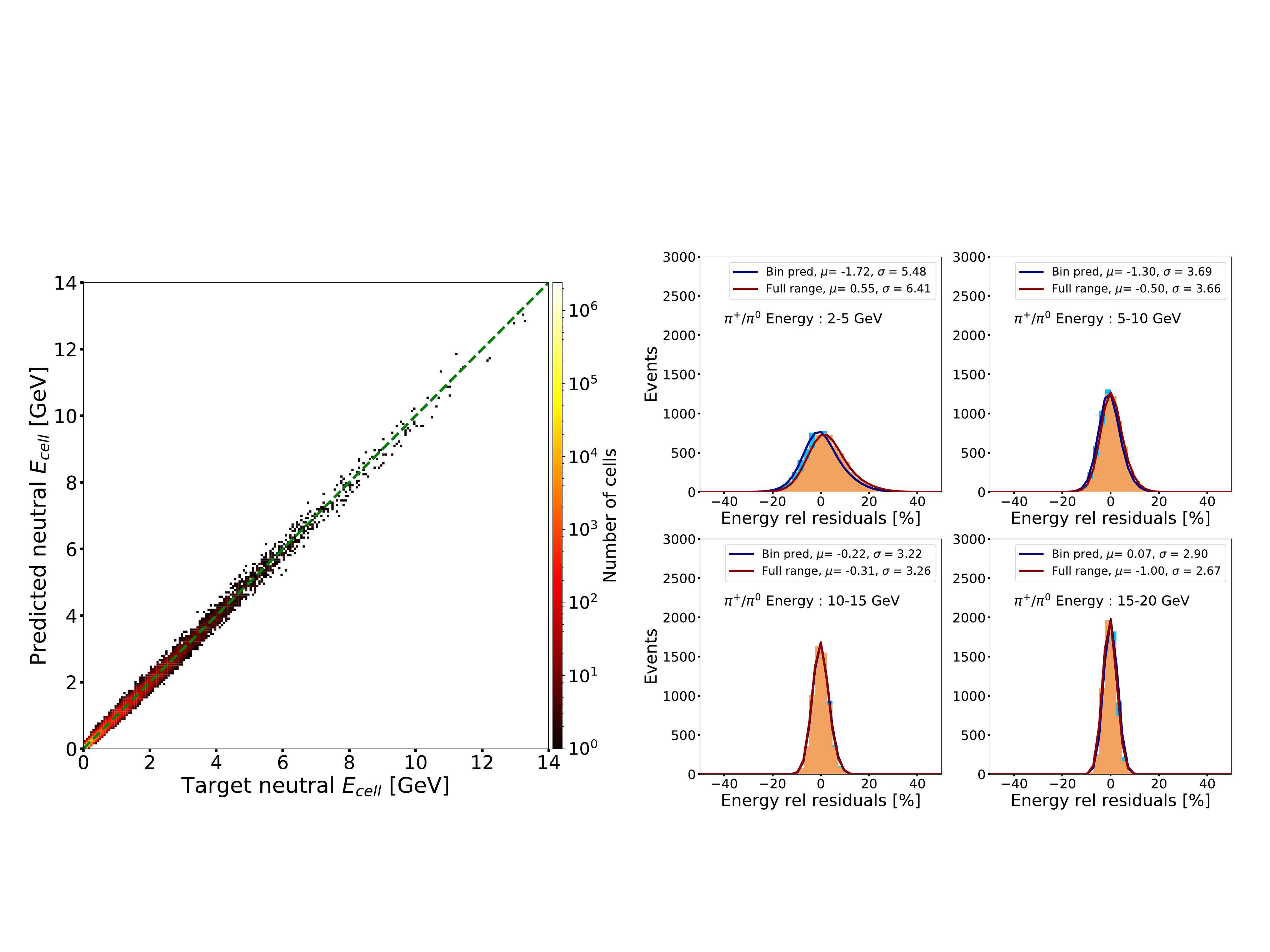}
 \end{minipage}
 \vspace{-2cm}
 \caption{(left) Distribution of the predicted cell neutral energy as a function of the truth deposited neutral energy for the entire energy range of interest. (right) Topological cluster level relative energy residuals 
 $(E_{predicted}-E_{Neutral})/(E_{Neutral})$ in specific energy ranges. The blue lines denote the performance from  trainings which are done on the specific energy ranges. The red curves show the performance of a unique training on a sample produced covering the entire energy range.}
  \label{Fig:comparison}
\end{figure*}

The Topological clustering algorithm groups cells  
based on their energies and topological location. 
The algorithm is designed to cluster cells originated 
by a single neutral or charged particle as well as 
to suppress noise.  The algorithm starts by ranking  
cells based on their significance over the corresponding nominal noise values.  Cells with a significance larger than five $\left( \frac{E}{\sigma} \geq 5\right)$ are considered as seeds and a topological search is performed on their adjacent cells in the longitudinal direction and on their adjacent and next-to adjacent cells in the transverse plane. 
If one of the clustered cells has a significance  
$\frac{E}{\sigma}$ $\geq$  2 , an additional clustering 
step is performed.  If a seed cell is found to be adjacent, within two cells to another topocluster, the two topoclusters are merged. The closest topocluster to the extrapolated track position is considered to be matched to the track.  

The expected charged energy is estimated using a 
parametrisation 
of the  energy deposited  
by a $\pi^{+}$ within the matched 
topocluster, referred to 
as  
$\langle E_{\text{pred}} \rangle$.
This parametrisation is 
computed from  
template distributions 
obtained  using a pure sample 
of $\pi^{+}$ without contamination 
of $\pi^{0}$ and it is dependent 
on the track momentum 
and the estimated Layer of 
First Interaction  (LFI, where the 
first nuclear interaction takes 
place).  The track position is  
extrapolated to the 
calorimeter layers and 
rings of radius 
equal to a single cell 
pitch are built.   
The rings are then ordered 
according to their expected energy density. The ring energies are
progressively subtracted from the topocluster in decreasing order of energy density. The algorithm 
proceeds  until the total amount of removed energy  
exceeds $\langle E_{\text{pred}} \rangle$.
 If the energy in the ring is larger than the required energy 
to reach  $ \langle E_{\text{pred}} \rangle$,  the energy in that ring is not 
fully subtracted but scaled to the fraction 
needed to reach the expected energy released by the $\pi^{+}$ .  
The remaining energy in the  topolcluster is considered as 
originating from neutral particles. 

\section{Results for standard resolution}
\label{sec:results}

An example event is displayed in Figure~\ref{Fig:EventDisplayML}, showing the truth energies and the predictions of the pPflow and ConvNet algorithms. It illustrates how this specific algorithm produces a more accurate image of the neutral energy deposit. The other algorithms that are considered provide very similar results. 
 

The training and the evaluation of the NN models are performed on cells belonging to the topoclusters. 
Figure~\ref{Fig:comparison} shows a comparison between the predicted and truth cell energies for the Graph model for an inclusive energy range of [2-20] GeV. It illustrates the ability of the ML models to predict the cell fractions over a wide range of energy. The distributions of residuals $\left(\frac{E_{predicted}-E_{Neutral}}{E_{Neutral}}\right)$ computed from the prediction of a network trained on specific energy ranges and a network trained over the entire energy range. Both, the predicted and the truth neutral energies used in the the residuals are computed as the sum of cell energies belonging to the topoclusters. The marginal differences observed show that the training performed over the full energy range provides very similar results compared to the specific model trained exclusively in each of energy regions.

To quantify the overall performance of the NN methods, two figures of merit related to the reconstruction of the neutral energy deposits are considered. 
The first is the energy resolution of the residual neutral energy after the charged energy subtraction. The second is the spatial resolution in the reconstruction of the location of the barycenter of the neutral energy in the second layer of the EM calorimeter.

\begin{figure*} 
\begin{minipage}{\columnwidth}

\includegraphics[width=2.2\textwidth]{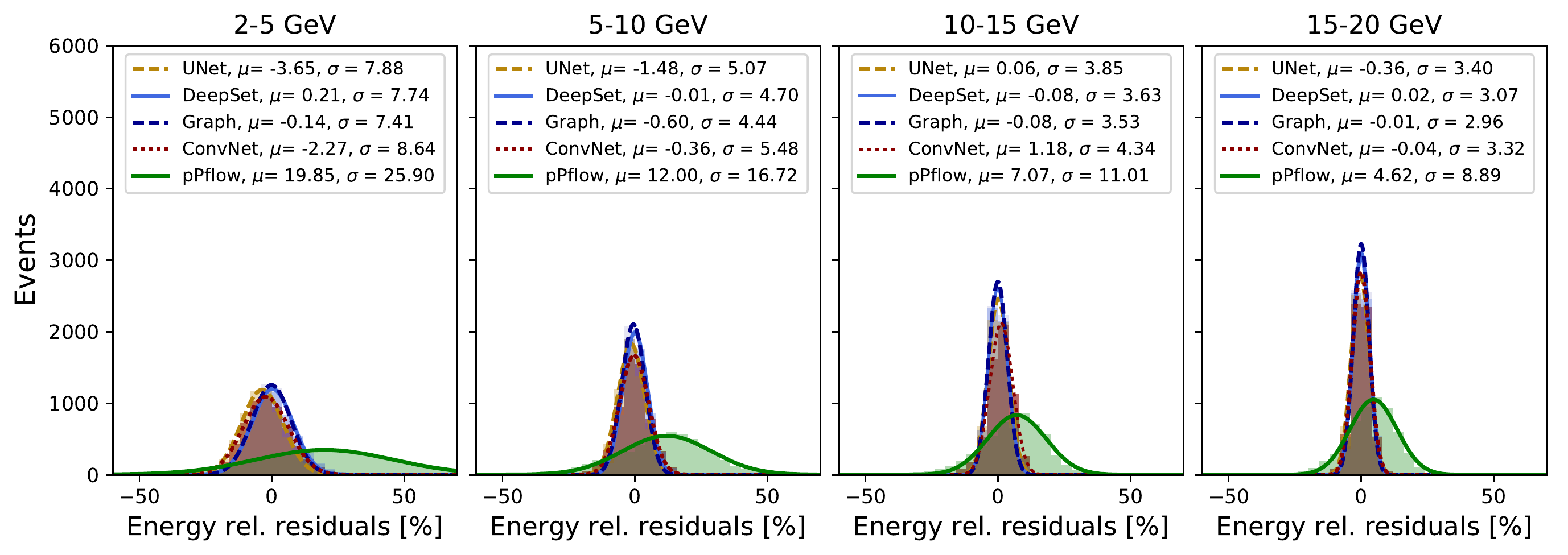}
 \end{minipage}
 \caption{The neutral energy relative residuals 
 $(E_{predicted}-E_{Neutral})/(E_{Neutral})$ distributions of the pPflow and the different NN algorithms for different energy ranges described in the text. The distributions are fit with a sum of two gaussians to catch non-gaussian tails. The values of the central gaussians are shown in the plots. In order to compare with pPflow performance, all the residuals are computed from the topocluster closest to the track.}
  \label{Fig:Perform2to5_E}
\end{figure*}

 
Figure \ref{Fig:Perform2to5_E} shows the neutral energy 
residual
distributions for the pPflow algorithm and all NN models. The cells considered to compute the residual are those pertaining to the initial topocluster. The distributions are fitted with a sum of two Gaussian distributions, to quantify the main gaussian resolution and the non-gaussian tails.  For all the NN models, non-gaussian tails are below 2\% for all energy ranges and all models.
The estimated energy resolutions  of 
the ML algorithms show an improvement in excess of a factor of 4 compared to the pPflow in the lowest energy range corresponding to 2-5 $GeV$. The relative improvement progressively decreases to reach approximately a factor of 2 in the highest energy range of 15-20 $GeV$. 

For the  pPflow algorithm a bias in the mean, ranging from 20\% in the low energy range, to 5~\% in the highest energy range, is observed. It is found to originate from the parametrisation of the charged particle energy deposits predictions which is derived from a pure sample of $\pi^{+}$. The size and energy of the topocluster in isolated $\pi^{+}$ are systematically lower compared to the evaluation sample which features a large  $\pi^{0}$ and $\pi^{+}$ overlap. As the energy of the $\pi^{+}$ increases, the size of the topoclusters in both, isolated and overlapped topologies becomes comparable thus reducing the bias in the high energy ranges. In contrast, for all NN methods and most energy ranges, the predictions yield a precise average value. In the low energy range, the UNet and CNN algorithms show small biases of approximately 4\% and 2\% respectively. These small biases are not observed for the GNN and Deepset algorithms. The GNN and DeepSet methods also perform up to approximately 20\% better in terms of neutral energy response resolution. While these differences could be due to the tuning of the model or specific trainings that could benefit from different training strategies or larger training datasets. These differences are sufficiently significant to be emphasised.

\begin{figure*} 
\begin{minipage}{\columnwidth}
\includegraphics[width=2.2\textwidth]{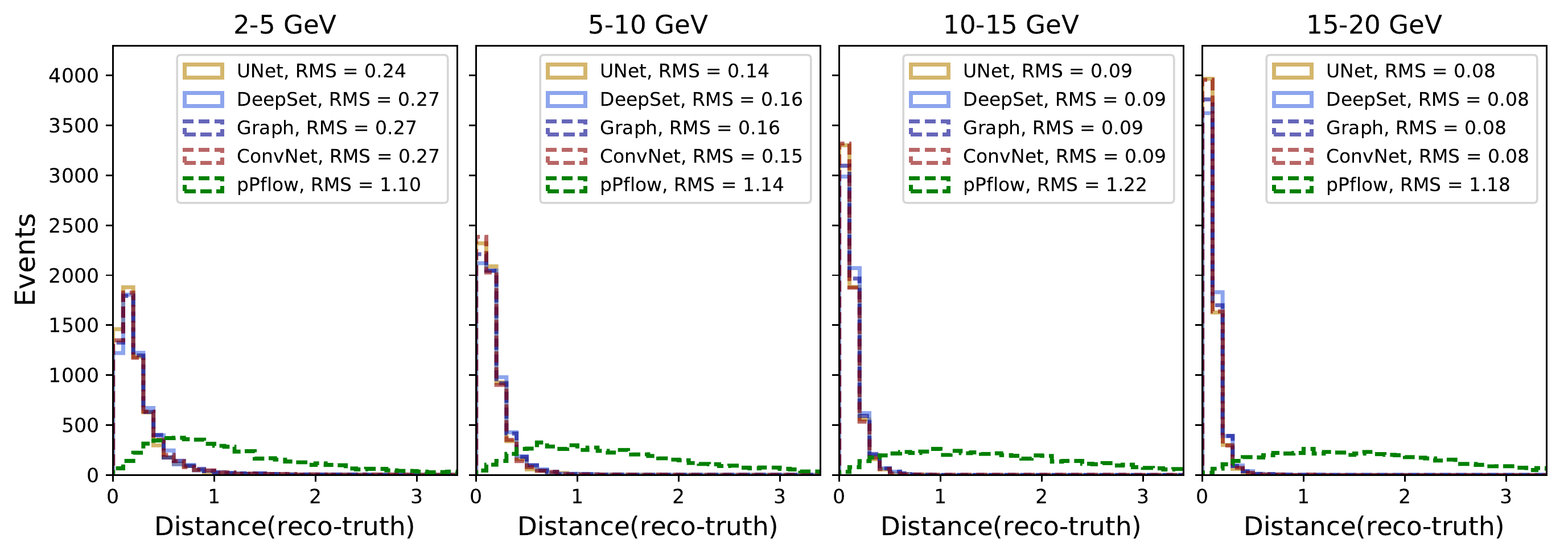}
 \end{minipage}
 \caption{The distance computed in number of cells
 between the barycenter of the predicted and truth
 neutral energy in the ECAL2 layer when using the
 pPflow or the NN algorithms  for different energy
 ranges described in the text. The equivalent
 distributions in  different calorimeter layers show
 a similar behaviour.}
  \label{Fig:Perform2to5_T}
\end{figure*}

Figure~\ref{Fig:Perform2to5_T} illustrates the distance, in number of cells, between the barycenter of the truth neutral energy  and the predicted neutral energy within the topocluster for the different algorithms in the ECAL2 layer. The NN models provide an accurate estimate of the barycenter, outperforming the pPflow results. While the RMS of the pPflow predictions is approximately constant as a function of the energy, the NNs predictions improve their accuracy as a function of the pions energy. The main reason for the improvement is related to the pPflow energy subtraction which is performed within rings around the extrapolated track position and therefore it gives a very approximate estimate of the precise topology of the neutral energy deposit, in contrast with the NN algorithms. 
Improvements of a factor of 4 or more in the spatial resolution are observed for the ML algorithms versus the pPflow algorithm. 

In the next Section, super-resolution models used to further improve the spatial resolution of the calorimeter system will be discussed. 

\section{Results for super-resolution}
\label{sec:superres}

 In HEP experiments very large samples, providing a precise description of the development of electromagnetic and hadronic showers, can be simulated.

Precise calorimeter images with arbitrarily high granularity can thus be produced, providing  high-resolution images that can be used as targets in the training of NN algorithms. We use such a dataset with information of low granularity data and high granularity truth information to quantitatively establish a proof-of-principle NN based super-resolution method.

The models used in
Section~\ref{sec:results} to predict
the fraction of cell energy pertaining
to neutral hadrons, in some cases
supplemented with a broadcasting layer
to perform changes in granularity, can
also be used to predict calorimeter
images with augmented granularity. To
efficiently showcase the ability of
these methods, the LG detector
configuration is used as a baseline.
the HG detector is then used as the
target of super-resolution algorithms.
To avoid overlapping photons in the
high-resolution image, only the sample
featuring  2-5~GeV pions was used for
these studies.

\begin{figure*}
\begin{minipage}{\columnwidth}
\hspace{1.8cm}
\includegraphics[width=1.6\textwidth]{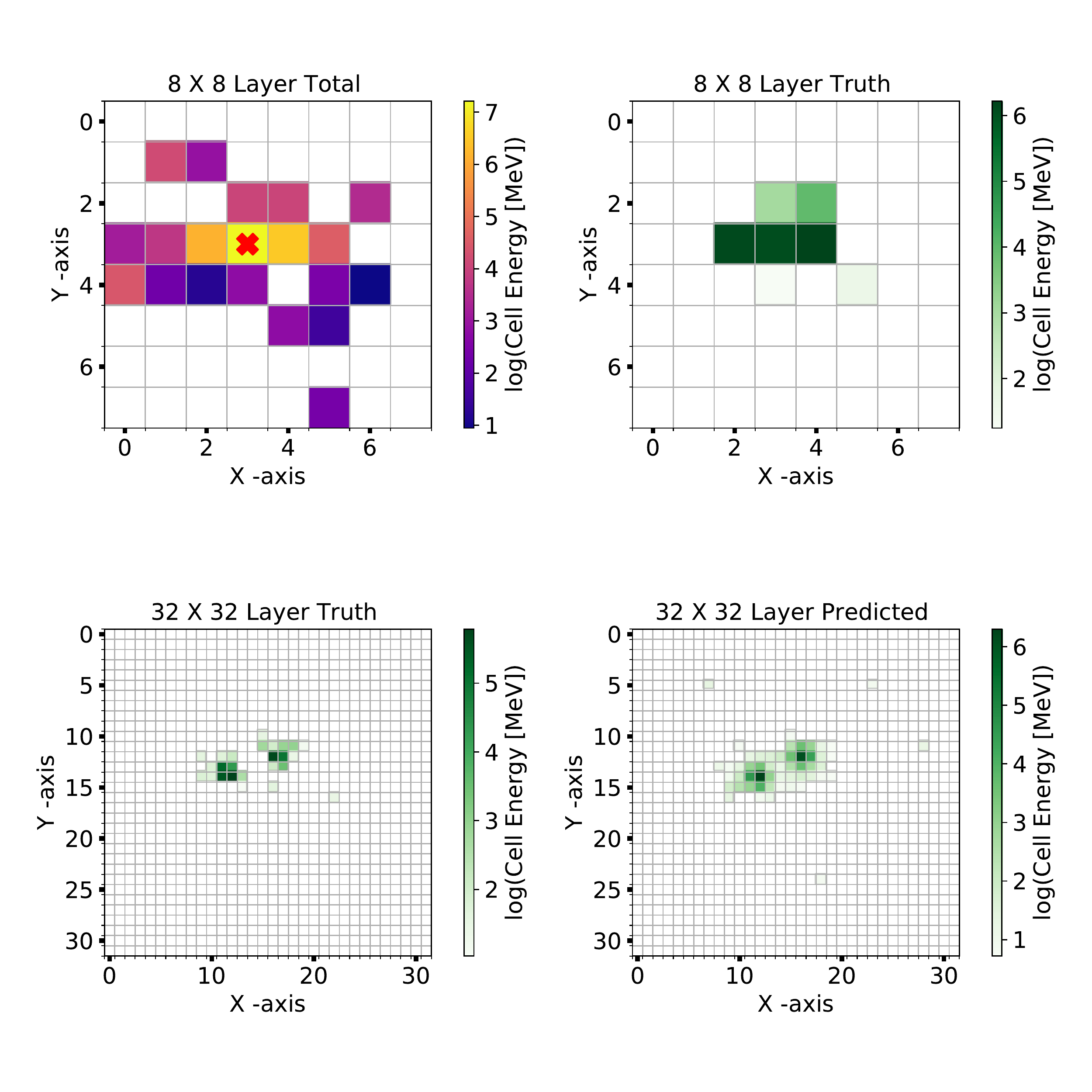}
 \end{minipage}
 \caption{An event display of total energy shower (within topocluster), as captured by a calorimeter layer of $8 \times 8$ granularity, along with the location of the track, denoted by a red cross (left) and the same shower is captured by a calorimeter layer of $32 \times 32$ granularity (middle). The bottom right panel shows the corresponding event predicted by the NN. The figure shows that the shower originating from a $\pi^{0} \rightarrow \gamma \gamma$ is resolved by a $32 \times 32$ granularity layer.}
 \label{Fig:Super_Res_EvDisplay}
\end{figure*}

An example event is displayed in Figure~\ref{Fig:Super_Res_EvDisplay}, illustrating the truth neutral energy in the second ECAL layer for the LG and the HG detector configurations along with the high-resolution predictions for the Graph+UNet algorithm. This example illustrates the ability of the NN, not only to subtract the correct amount of energy from the charged particle, but also to provide an accurate and higher granularity prediction of the energy deposited by the two photons which can be seen to be separated and to nicely reproduce the underlying true pattern of energy deposition. 

\begin{figure*}
\begin{minipage}{\columnwidth }
\hspace{1.0cm}
\includegraphics[width=1.8\textwidth, trim=0cm 3.0cm 0cm 3cm, clip=true]{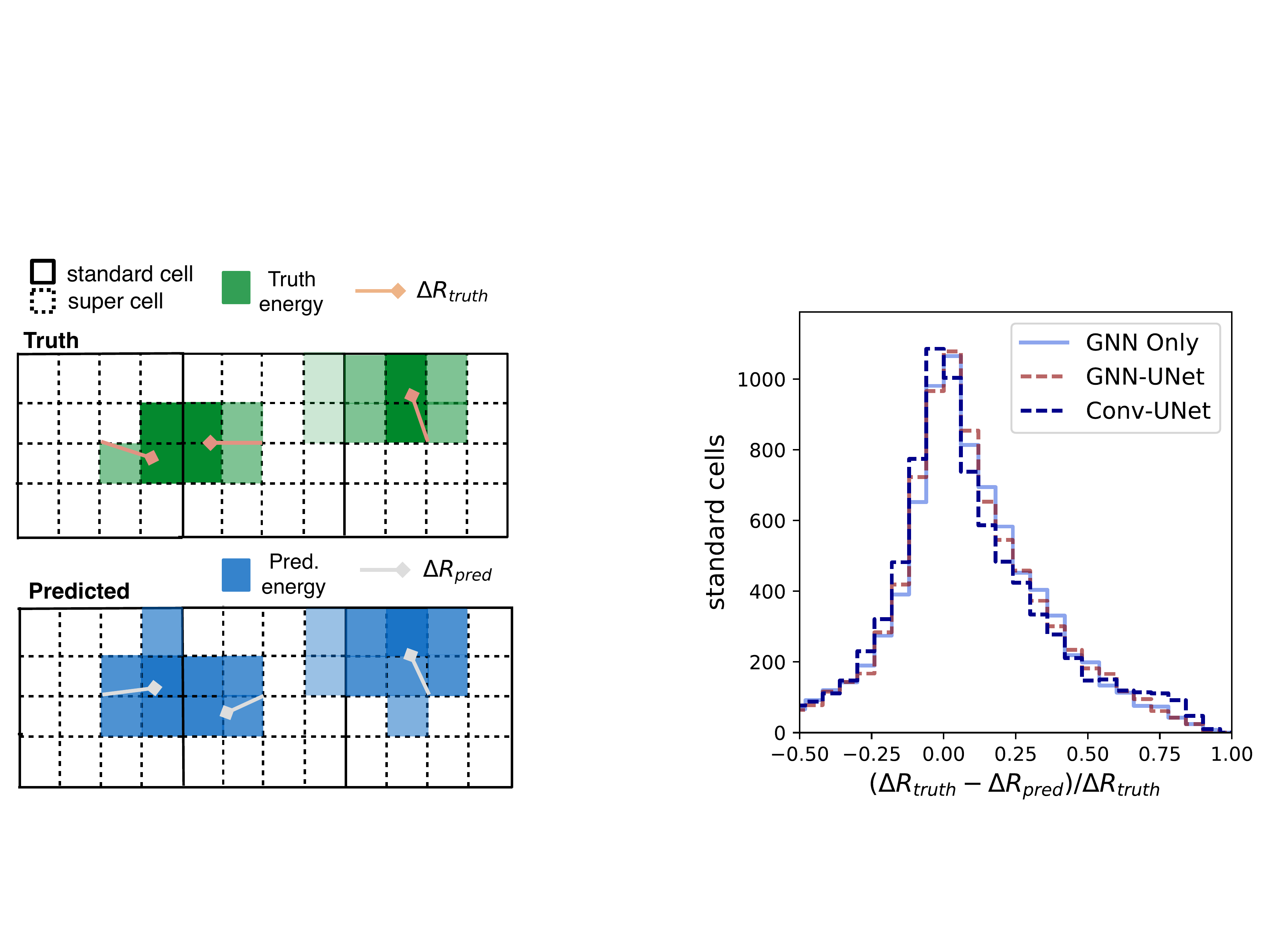}
 \end{minipage}
 \caption{(left) Schematic representation of three cells of the LG detector and the respective high-resolution $super$ cells. The truth and predicted neutral energy component in the high resolution image are outlined in green and blue, respectively.
 The radial distance ($\Delta \textrm{R}$) between the barycenter of the neutral energy distribution of the super cells and the center of the corresponding standard cell is outlined in blue. (right) Relative residuals distributions of the radial distance for the different NN architectures.}
 \label{Fig:dRPlotNew}
\end{figure*}

To quantify the performance of the super-resolution algorithms, the distance of the super-resolution energy barycenter to the center of the initial granularity cell is calculated for both the true energy deposits and the prediction, as illustrated in Figure~\ref{Fig:dRPlotNew}. The relative difference between these distances ($\Delta \textrm{R}$) is also shown in Figure~\ref{Fig:dRPlotNew} for all methods, thus showing that the super-resolution methods are able to reproduce the correct barycenter distance to within a good precision. The different NN models show similar results.

\begin{figure*}
\begin{minipage}{\columnwidth }
\includegraphics[width=2.\textwidth,  trim=0cm 7.0cm 0cm 6cm, clip=true]{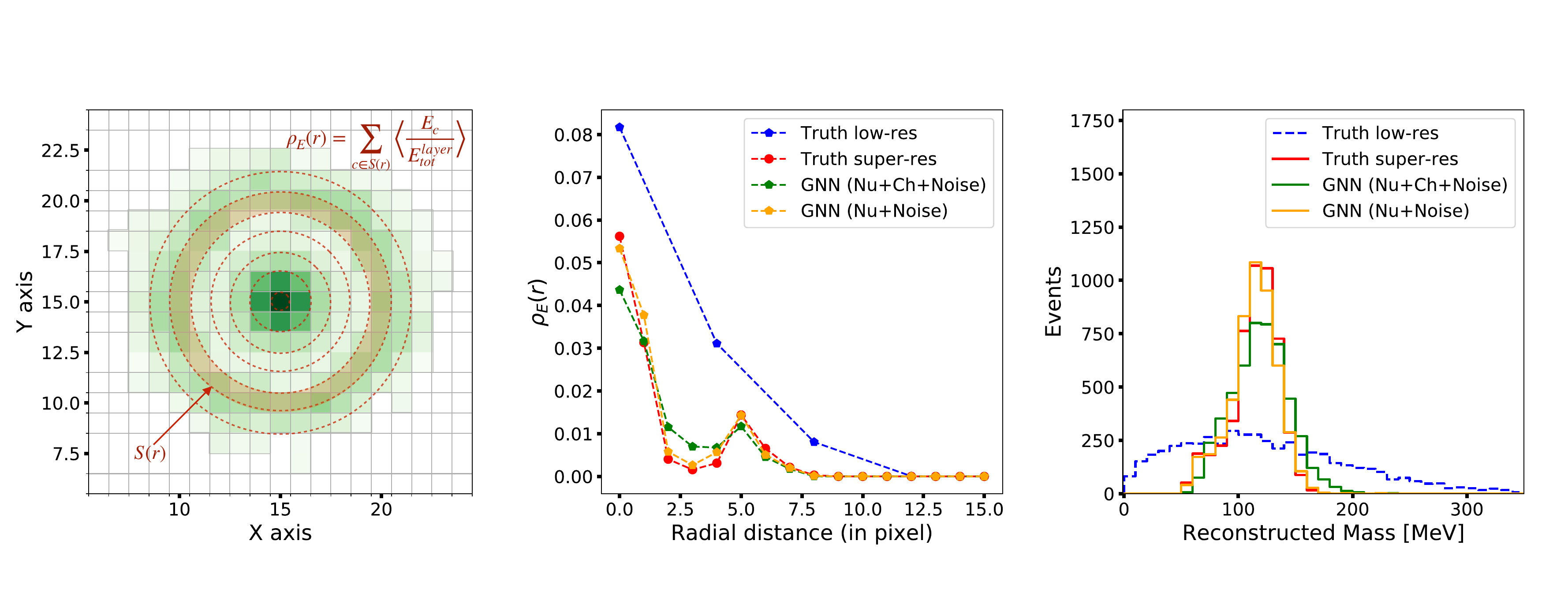}
\end{minipage}
 \caption{
 (left) Average cell truth neutral energy normalised to the total energy in the second layer of the HG calorimeter, where calorimeter images have been systematically centred on the highest energy cell.  From this distribution the average energy profile as a function of radial distance $\rho_{E}(r)$ is computed (as displayed in the left panel) from this 2-D distribution in rings (of surface $S(r)$) with a width of one cell (middle). (right) Calibrated reconstructed mass distribution from the neutral energy calorimeter in the LG case in blue, the truth HG case in red and the super resolution predicted HG image in green and orange. The green curve is the prediction when there is charged and neutral energy overlapping along with noise. The orange curve, which follows closely the truth distribution, shows the prediction when a training and evaluation is carried out on sample consisting of neutral energy and noise only. This demonstrates the degradation originates from the presence of overlapping charged shower. All predictions are computed from the GNN algorithm. The other approaches considered herein give similar results.}
 \label{Fig:Mass_Profile}
\end{figure*}

To illustrate the performance of the super-resolution methods in producing an augmented granularity calorimeter image,  
 an average image is produced from a large sample where all images are centred on the highest energy cell, as illustrated in the left panel of Figure \ref{Fig:Mass_Profile}. The appearance of a circular secondary maximum energy ring around the central highest energy spot denotes the presence of a second photon from a neutral pion decay.  
The same figure (middle panel), shows that the integrated average relative energy profile $\rho_{E}(r)$,  obtained from the
low-resolution image (blue line) cannot capture the secondary peak, while the super-resolution predictions are able to capture the secondary peak (green). The secondary peak location from the prediction coincides with the position of the peak in the true neutral particle energy distribution (red); however, the predicted energy distribution displays a slight underestimate under the primary peak and an overestimate between the two peaks. This would result in a degradation of the discrimination between a photon and a $\pi^0$ at reconstruction level. To further check the origin of this degradation, a super-resolution GNN is trained on a sample without an overlapping charged pion. In this case the super-resolution prediction (orange) reproduces accurately the expected truth energy density distribution (red). The degradation is relatively minor given the large overlap between the charged and neutral pions imposed in the nominal low energy (2-5~GeV) sample chosen here, where typically the angular distance between the charged and neutral pions is smaller than the opening angle of the two photons from the $\pi^0$ decay. 


Another illustration of the performance of the ability of high-resolution layers in resolving the two photons clusters
is obtained by reconstructing the invariant mass
 of the $\pi^{0}$ from the energies of the two photons. The individual photon energies and directions are estimated using a $k$-mean algorithm applied to the
predicted neutral energy with a
number of clusters equal to two ($k=2$). To better illustrate the impact on the mass resolution, the reconstructed photon energies are calibrated in order to yield in all cases the $\pi^0$ mass. The rightmost panel in Figure~\ref{Fig:Mass_Profile}
shows a comparison between the reconstructed mass
distribution for low- and high-resolution layers and illustrates how the super-resolution GNN predictions are able to produce a mass peak with a resolution close to that of the native high-resolution images.

For the HG calorimeter, the spatial resolution of the shower allows to capture two distinct photon energy clusters using the $k$-mean algorithm. For the LG detector, the spatial resolution is missing and hence the $k$-mean algorithm fails to identify two peaks distinctively. Hence when we try to construct an invariant mass spectrum from two reconstructed clusters, 
the output of HG detector gives a well reconstructed peak whereas a relatively flat distribution is obtained from the LG shower.

\section{Conclusion and outlook}
\label{sec:conclusions}

Particle Flow reconstruction has an important role in high energy particle collider experiments and is being considered for the design of future experiments. 
A key component of particle flow reconstruction is the ability to distinguish neutral from charged energy deposits in the calorimeters.
In this paper, a Computer Vision approach to this task, based on calorimeter images,
is proposed. This approach explores the ability of Deep Learning techniques to produce calorimeter images of the energy deposits with an optimal separation of the energy deposits originating from neutral particles from those originating from charged particles. Several schemes based on Convolutional layers with the insertion of up-convolution and down-convolution blocks, dynamic graph convolution 
networks and DeepSet networks are proposed.  \\

The detailed performance is quantified using a simplified layout of electromagnetic and hadron calorimeters, focusing on the challenging case of overlapping showers of charged and neutral pions. An improved energy and direction reconstruction of the initial particles is obtained, compared to parametric models. Enhanced calorimeter images of the event with a finer granularity with respect to the apparatus' native granularity are also obtained using super-resolution techniques. All the techniques used yield excellent performances at producing calorimeter images. The GNN and DeepSet approaches; however, appear to provide a slightly better resolution and more stable results over a wide range of energies. These algorithms constitute an improved first step towards a computer vision Particle Flow algorithm or a powerful intermediate step in the development of precise identification algorithms.

\indent

\section{Acknowledgements}
SG, EG and JS were supported by the NSF-BSF Grant \\ 
2017600,
the ISF Grant 2871/19 and the Benoziyo Center for High Energy Physics.

\bibliographystyle{unsrt}
\bibliography{PFlowDocument}{}

\appendix

\section{Up and Down Convolution description}
\label{sec:upsample}

The UpConv block is built out of two-dimensional convolutional layers (Conv2D) with 
$3 \times 3$ kernel size, followed by batch-normalization (BN) layers and 
Leaky ReLU activation function  \cite{xu2015empirical}, with slope parameter $\beta=0.1$. 
A single UpConv
block consists of five sequences of Conv2D, BN and Leaky ReLU activation function
followed by pixel shuffle upsampling~\cite{shi2016realtime} operation.
The pixel shuffle process converts a 
multi-channel low-resolution image to a higher
resolution image with a lower number of channels 
such that the total number of pixels in both
the images are equal. One such NN block is 
presented in Equation~\ref{eqn:UpConvFlow}.
The $\rm Pix\_Shuf(u\_f)$ in the same equation 
represents a pixel shuffle layer with 
upscale factor $\rm u_f$. 

The DownConv blocks are made out of Conv2D layers whose kernel size and stride depend on the down-scaling factor. For example if we want to reduce the resolution of an image from 
size $64 \times 64$ to $16 \times 16$, i.e. a 
reduction by a scale factor 4, we use 
a layer Conv2D[1, 1, K=(4,4), stride=(4,4)].
\begin{equation}
  \begin{aligned}
    L_{N \times N} & \rightarrow Conv2D\Big(1, 16, K = (3,3) \Big) \rightarrow LR(\beta=0.1) + BN\\
      & \rightarrow Conv2D\Big(16, 32, K = (3,3) \Big) \rightarrow LR(\beta=0.1) + BN\\
      & \rightarrow Conv2D\Big(32, 64, K = (3,3) \Big) \rightarrow LR(\beta=0.1) + BN\\
      & \rightarrow Conv2D\Big(64, 128, K = (3,3) \Big) \rightarrow LR(\beta=0.1) + 
      BN \\
      &  \rightarrow Conv2D\Big(128, 256, K = (3,3) \Big) \rightarrow LR(\beta=0.1) + 
      BN \\
      & \rightarrow Pix\_Shuf(u\_f) \\
      &  \rightarrow Conv2D\Big(256/(u\_f)^2, 1, K = (1,1) \Big) \\
      & \rightarrow ReLU \rightarrow 
      L_{64 \times 64} \\
      & {\rm where,} ~ 
      u\_f \times N = 64.
  \end{aligned}
  \label{eqn:UpConvFlow}
\end{equation}

\section{Model descriptions of ConvNet, GraphNet and Deepset}
\label{sec:graphim}
\begin{figure*} 
\begin{minipage}{\columnwidth}
\includegraphics[width=2\textwidth]{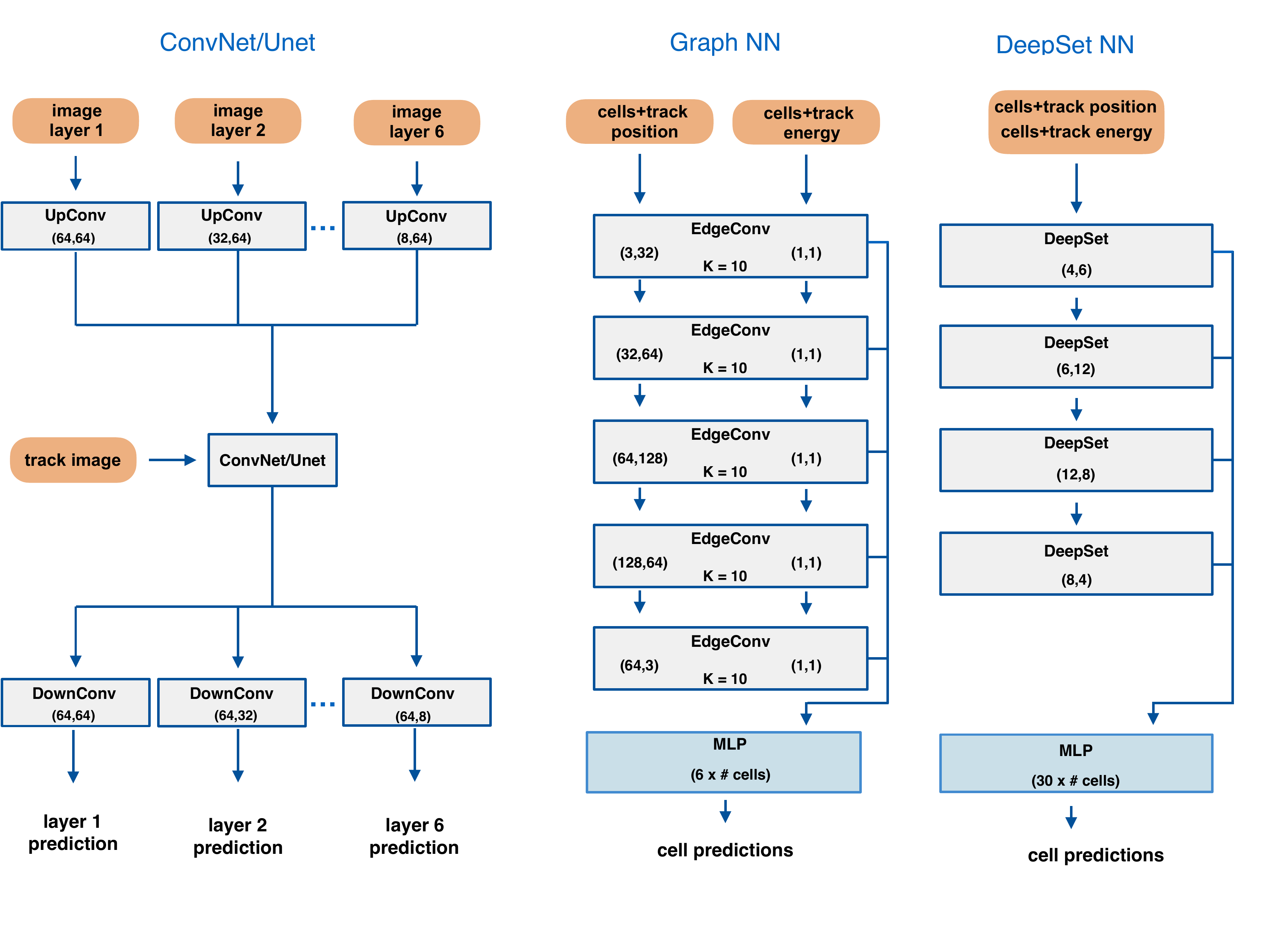}
\end{minipage}
\caption{An illustration of the general convolutional network with the UNet variant is shown on the left. 
In the middle the general flow of the graph neural network with dynamic EdgeConvolution layers is illustrated. Each cells and the track is treated as a node, which are passed through EdgeConv layers. Finally output energy representation of all the layers are processed through a MLP to predict neutral energy fraction for the individual cells. 
An illustration of the deepset network model (on the right) built out of several deepset layers. 
}
\label{Fig:ECModel}
\end{figure*}

The generalized layouts of convolutional network, including UNet, Graph network and Deepset network, which are described in details in section \ref{sec:model}, are shown in Figure \ref{Fig:ECModel}. \\ The left panel in the figure shows a general layout of the image based convolutional model. Different input layers with varying granularities are brought to equal footing with appropriate upconvolution layers. The track image is added as an additional layer. The seven channel image is then mapped to a six channel image through a combination of convolutional network or UNet. Then appropriate down convolutional layers are applied to make the output images as same granularity to that of input. The description of this model is discussed in further details in section \ref{sec:convnetmodel}.

The middle panel shows the general layout for the graph network, using dynamic EdgeConvolutional layers. The network is described in details in \ref{sec:graphnet} and comprises of 
five message passing layers referred to as {\it "EdgeConv"}. Each layer consists
of four multi-layer perceptrons (MLP) denoted $\Theta_{x}$ and $\Phi_{x}$ which act only on spatial co-ordinates {\it x}, and $\Theta_{e}$ and $\Phi_{e}$ which act on energies {\it e} only. The graphs are dynamically defined based on the K-nearest neighbors algorithm (K-NN) using K=10. 
Here the K-NN algorithm acts only on the co-ordinates {\it x} to 
dynamically construct the graph.
The dimensionality of each node before and after the message passing are denoted $M$ and $N$ respectively. 

The message passing operation, referred to as  
{\it "EdgeConv"}, from layer $l$ to layer $l+1$ consists in the following operations: 
\begin{equation*}
 \begin{aligned}
 x_{i}^{l+1} = max_{j \in \mathcal{N}(i)} ~  \{ \Theta_{x} ( x_{j}^l - x_{i}^l ) \} + \Phi_{x} ( x_{i}^l ), \\
 e_{i}^{l+1} = mean_{j \in \mathcal{N}(i)} ~ \{ \Theta_{e} ( e_{j}^l - e_{i}^l ) \} + \Phi_{e}  (e_{i}^l),
 \end{aligned}
 \label{eqn:GraphConv}
\end{equation*}

\noindent where $\mathcal{N}(i)$ is the set of  indices $j$ of the $K$ neighboring nodes of any given node $i$. 
For the spatial coordinates the ($M ~\,, ~N$) 
values chosen are (3, 32), (32, 64), (64, 128), (128, 64), 
(64, 3). For the energies, the input and output dimensions are 1 for all layers.

\end{document}